\documentclass[twocolumn,preprintnumbers,amsmath,amssymb,aps,pra]{revtex4-1}

\usepackage[german,english]{babel}
\usepackage[utf8]{inputenc}
\usepackage{graphicx}
\usepackage{mathtools}

\usepackage{color}

\usepackage{dcolumn}
\newcolumntype{d}{D{.}{.}{3}}

\newcommand*{\coord}[1]{\mathbf{#1}}

\newcommand*{\vecr}{\coord{r}}

\newcommand*{\vecx}{\coord{x}}

\newcommand*{\binteg}[3]{\int^{\mathrlap{#3}}_{\mathrlap{#2}}\ud{#1}\,}
\newcommand*{\integ}[1]{\int\!\!\!\:\ud{#1}\:}

\newcommand*{\crea}[1]{\hat{#1}^{\dagger}}
\newcommand*{\anni}[1]{\hat{#1}^{\vphantom{\dagger}}}

\newcommand{\brakket}[3]{\langle{#1}|{#2}|{#3}\rangle}

\DeclareMathOperator{\BesselI}{\mathrm{I}}
\DeclareMathOperator{\BesselJ}{\mathrm{J}}

\DeclareMathOperator{\LegendreP}{\mathrm{P}}
\DeclareMathOperator{\Order}{\mathcal{O}}

\DeclareMathOperator{\sgn}{sgn}

\newcommand*{\abs}[1]{\lvert#1\rvert}

\newcommand*{\du}{\partial}
\newcommand*{\e}{\textrm{e}}

\newcommand*{\half}{\frac{1}{2}}

\newcommand*{\isDefinedAs}{\coloneqq}
\newcommand*{\Nats}{\mathbb{N}}

\newcommand*{\ud}{\mathrm{d}}

\allowdisplaybreaks[4]

\usepackage{hyperref}

\begin{document}

\title{Long-range interactions and the sign of natural amplitudes in two-electron systems}
\author{Klaas J. H. Giesbertz}
\affiliation{Theoretical Chemistry, Faculty of Exact Sciences, VU University, De Boelelaan 1083, 1081 HV Amsterdam, The Netherlands}
\author{Robert van Leeuwen}
\affiliation{Department of Physics, Nanoscience Center, University of Jyväskylä, P.O. Box 35, 40014 Jyväskylä, Survontie 9, Jyväskylä, Finland}

\date{\today}

\begin{abstract}
In singlet two-electron systems the natural occupation numbers of the one-particle reduced density matrix are given as squares of the natural amplitudes which are defined as the expansion coefficients of the two-electron wave function in a natural orbital basis. In this work we relate the sign of the natural amplitudes to the nature of the two-body interaction. We show that long-range Coulomb-type interactions are responsible for the appearance of positive amplitudes and give both analytical and numerical examples that illustrate how the long-distance structure of the wave function affects these amplitudes. We further demonstrate that the amplitudes show an avoided crossing behavior as function of a parameter in the Hamiltonian and use this feature to show that these amplitudes never become zero, except for special interactions in which infinitely many of them can become zero simultaneously when changing the interaction strength. This mechanism of avoided crossings provides an alternative argument for the non-vanishing of the natural occupation numbers
in Coulomb systems.
\end{abstract}

\maketitle

\section{Introduction}

Recently, the question if the natural occupation numbers can become zero has regained interest~\cite{CioslowskiPernal2000b, CioslowskiPernal2006, ShengMentelGritsenko2013, GiesbertzLeeuwen2013a}. The natural occupation numbers are defined to be the eigenvalues of the one-body reduced density matrix (1RDM), which is given in second quantization as
\begin{align*}
\gamma(\vecx,\vecx') \isDefinedAs \brakket{\Psi}{\crea{\psi}(\vecx')\anni{\psi}(\vecx)}{\Psi}
= \sum_kn_k\phi_k(\vecx)\phi^*_k(\vecx'),
\end{align*}
where $\vecx \isDefinedAs \vecr\sigma$ denotes a combined space and spin coordinate. The corresponding eigenfunctions are called the natural orbitals (NOs).

The question whether the occupation numbers vanish is interesting for a number of applications. For the usual ab-initio methods that try to approximate the full many-body wavefunction by assembling linear combinations of Slater determinants constructed out of an orbital basis, the existence of only a finite number of (fractionally) occupied NOs implies that only these orbitals need to be included in the orbital basis to represent the wavefunction exactly~\cite{Lowdin1955a}. Such a situtation would be beneficial for the ab-initio methods, since extrapolations to the complete basis would not be required anymore. For the foundations of 1RDM functional theory~\cite{Gilbert1975} and the extended Koopmans' theorem~\cite{SmithDay1975, MorrellParrLevy1975}, however, the existence of vanishing occupation numbers poses some problems. In the case of 1RDM functional theory, the mapping between the 1RDM and the corresponding non-local potential becomes less unique, which makes it more difficult to properly define an inverse of the mapping $v_{\text{non-loc}} \mapsto \gamma$, cf.\ the invertibility of the potential-density mapping, $v_{\text{loc}} \mapsto \rho$, in density functional theory. The possible lack of invertibility of $v_{\text{non-loc}} \mapsto \gamma$ poses some difficulties in the formal foundations of 1RDM functional theory~\cite{PhD-Giesbertz2010}. For the extended Koopmans' theorem vanishing occupation numbers are probably more problematic~\cite{MorrisonAyers1995}. The extended Koopmans' theorem provides a method to calculate the ionization potentials from any approximation to the exact many-body wavefunction. It guarantees that the exact ionization potentials are obtained provided the set of (fractionally) occupied NOs is complete, i.e.\ no NOs exist which have an occupation number exactly equal to zero. In the case some of the natural occupations vanish, the ionization potentials from the extended Koopmans' theorem are not necessarily exact anymore.

The question whether the occupation numbers can be zero is not only complicated due to the construction of the solution to the Schrödinger equation, but additionally, many features of the wavefunction are smoothened by the integration to obtain the 1RDM. For example the Coulomb interaction requires the wavefunction to have a cusp at the coalescence points of the electrons, $\Psi(r_{12} \to 0) \sim 1 + \half r_{12}$~\cite{Kato1957, PackBrown1966}, so the wavefunction is discontinuous in its first derivative at these points. The inter-electronic cusp in the wavefunction also introduces a discontinuity in the 1RDM if the two arguments are close together, though, since we take effectively the square of the wavefunction and due to the volume element, the discontinuity is only $\gamma(\abs{r-r'} \to 0) \sim \abs{r-r'}^4$~\cite{Kimball1975}. In general smoothening of the cusp increases the decay rate of the natural occupation numbers according to Weyl's theorem~\cite{GiesbertzLeeuwen2013a}. To avoid this additional complication all studies of finite systems have focussed on singlet two-electron systems, since the NOs and their occupation numbers can directly be calculated from the wavefunction itself as follows. Because the spatial part of the wavefunction is symmetric, it can by diagonalized as
\begin{align}\label{eq:PsiSpectral}
\Psi(\vecr_1,\vecr_2) = \sum_kc_k\phi_k(\vecr_1)\phi_k(\vecr_2).
\end{align}
By constructing the 1RDM, one readily finds that the eigenfunctions of the wavefunction $\phi_k(\vecx)$ are also NOs and that the eigenvalues of the wavefunction are related to the natural occupation numbers as $n_k = c_k^2$. Hence, in the case of singlet two-electron systems, the additional integration to obtain the 1RDM can be avoided and the behavior of the NOs and their occupation numbers can directly be related to features of the wavefunction. Note that the 1RDM contains almost the same amount of information as the wavefunction, except for the sign of the coefficients, so the coefficients can vary over a larger range $-1 \leq c_k \leq 1$. The coefficients in the NO expansion of the two-electron wavefunction are also known as the natural amplitudes~\cite{BuijseBaerends2002}.

Cioslowski and Pernal used this feature to argue that the occupation numbers in the harmonium atom~\cite{CioslowskiPernal2000b} and the hydrogen molecule~\cite{CioslowskiPernal2006} can become zero. Both systems have a system parameter that allows one to change smoothly from the weakly correlated regime to the strongly correlated regime. The Hamiltonian of the harmonium is the identical to the one for the helium atom, except that the Coulomb potential of the nucleus has been replaced by a harmonic confinement, $\half\omega^2r^2$. By varying the strength of the harmonic potential, $\omega$, the system can be tuned in and out of the strong correlation regime. A high value of $\omega$ forces the electrons to be close to the nucleus, so this is the weakly correlated regime. For low values of $\omega$ the electrons are not forced to be close to the nucleus anymore, so they can avoid each other completely and the electrons becomes strongly correlated. For the hydrogen molecule this system parameter is the distance between the two hydrogen atoms. If the atoms are close together, the electrons are also necessarily close to each other to maximally benefit from the attractive potential of both nuclei. If the distance between the atoms is large, this benefit is lost and the electrons try to avoid each other maximally, so either electron 1 is on nucleus $a$ and electron 2 on nucleus $b$ or visa versa.

When making the transition from the weakly correlated regime to the strongly correlated regime, the signs of the most significant natural amplitudes show similar behavior in both systems. In the weakly correlated regime, typical numerical calculations give only one large positive natural amplitude. All other amplitudes are negative and much smaller in magnitude. In the strongly correlated regime the most significant natural amplitudes show an alternating pattern. The pattern of alternating signs has been explained by Cioslowski and Pernal for the hydrogen molecule by performing a perturbation analysis using the dissociation limit as the reference state~\cite{HirschfelderLowdin1959, HirschfelderLowdin1965, Davidson1976, CioslowskiPernal2006}. They found that the alternating sign pattern can be explained from stabilizing multipole terms in the perturbation expansion. The connection to the Van der Waals forces was readily made by Sheng et al.\ by analyzing the NO structure of the 2RDM for dissociating H$_2$~\cite{ShengMentelGritsenko2013}. The 2RDM clearly exhibits the characteristic Van de Waals multipole structure in the NO representation and these Van der Waals terms only lower the energy if the NO coefficients have an alternating sign pattern. The alternating sign pattern can therefore be regarded as a Van der Waals effect.

Cioslowski and Pernal used this observation to argue that the occupation numbers can vanish in both systems, since the natural amplitudes need to cross zero to change their sign. A difficulty in their reasoning is the assumption that the large amplitudes which differ in sign correspond to the \emph{same} NO. Indeed, a more careful inspection of the results for the harmonium atom by Cioslowski and Pernal~\cite{CioslowskiPernal2000b, GiesbertzLeeuwen2013a} clearly shows that the natural amplitudes which differ in sign correspond to different NOs. This observation is in agreement with our recent proof that the amplitudes in the harmonium atom can not become zero~\cite{GiesbertzLeeuwen2013a}. In the case of the hydrogen molecule Cioloslowski and Pernal used an additional argument by Goedecker and Umrigar in Ref.~\cite{GoedeckerUmrigar2000} where they argued that the helium atom should have only one positive NO coefficient under the assumption that the NOs are similar to the HF orbitals. Since there is only one positive natural amplitude in the united (helium) limit compared to dissociation limit, the coefficients of the NOs need to change sign along the bond stretching somewhere. They also showed numerical results where the coefficients of anti-bonding NOs crossing zero (Fig.~1 of Ref.~\cite{CioslowskiPernal2006}) corroborating their argument. However, there is always an infinite amount of weakly occupied NOs to take into account which is hard to cover with a finite basis representation of only cc-pV5Z quality. A more careful investigation by Sheng et al.\ showed that the zero-crossing actually disappears when adding more diffuse functions~\cite{ShengMentelGritsenko2013}. Further, they have pointed out that the assumption that the NOs are similar to the HF orbitals does not hold and they showed numerically that the helium atom actually does have more than one positive natural amplitude. Hence, the arguments of Cioslowski and Pernal that the natural amplitudes need to cross zero when the hydrogen bond is stretched do not hold and it is likely that the natural amplitudes actually never become zero.

There are still a number of open questions we would like to address in this article. Although it has been demonstrated that the NO coefficients do not cross zero, a clear explanation how the change in sign pattern is actually achieved by making the transition from the weakly to the strongly correlated regime, is still lacking. Before we can answer this question, we will first explain which features of the wavefunction cause the existence of positive natural amplitudes. Although the positive amplitudes have been regarded as a Van der Waals effect, they are also present in the helium atom~\cite{ShengMentelGritsenko2013}. Hence, the existence of multiple positive NO coefficients is due to a more fundamental property of the Coulomb interaction as will be exposed in the next section.

\section{Why positive natural amplitudes?}
\label{sec:posCoefs}

In this section we will address the question why there are always multiple positive natural amplitudes in Coulomb systems. A typical feature of Coulomb systems is that the wavefunction is required to have a cusp at the coalescence points of the electrons. However, if we calculate the NO coefficients of a simple Hylleraas wavefunction for a model atom in 1D of the form
\begin{align*}
\Psi(x_1,x_2) = K\alpha(x_1)\alpha(x_2)\bigl(1 + \eta\abs{x_1 - x_2}\bigr),
\end{align*}
where $K$ is a normalization constant and $\alpha(x) = \e^{-Z\abs{x}}$, we find that only one amplitude is positive and all the others negative~\cite{GiesbertzLeeuwen2013a}. Hence, it is not the cusp that causes the natural amplitudes to be positive.

\begin{figure}[t]
  \includegraphics[width=\columnwidth]{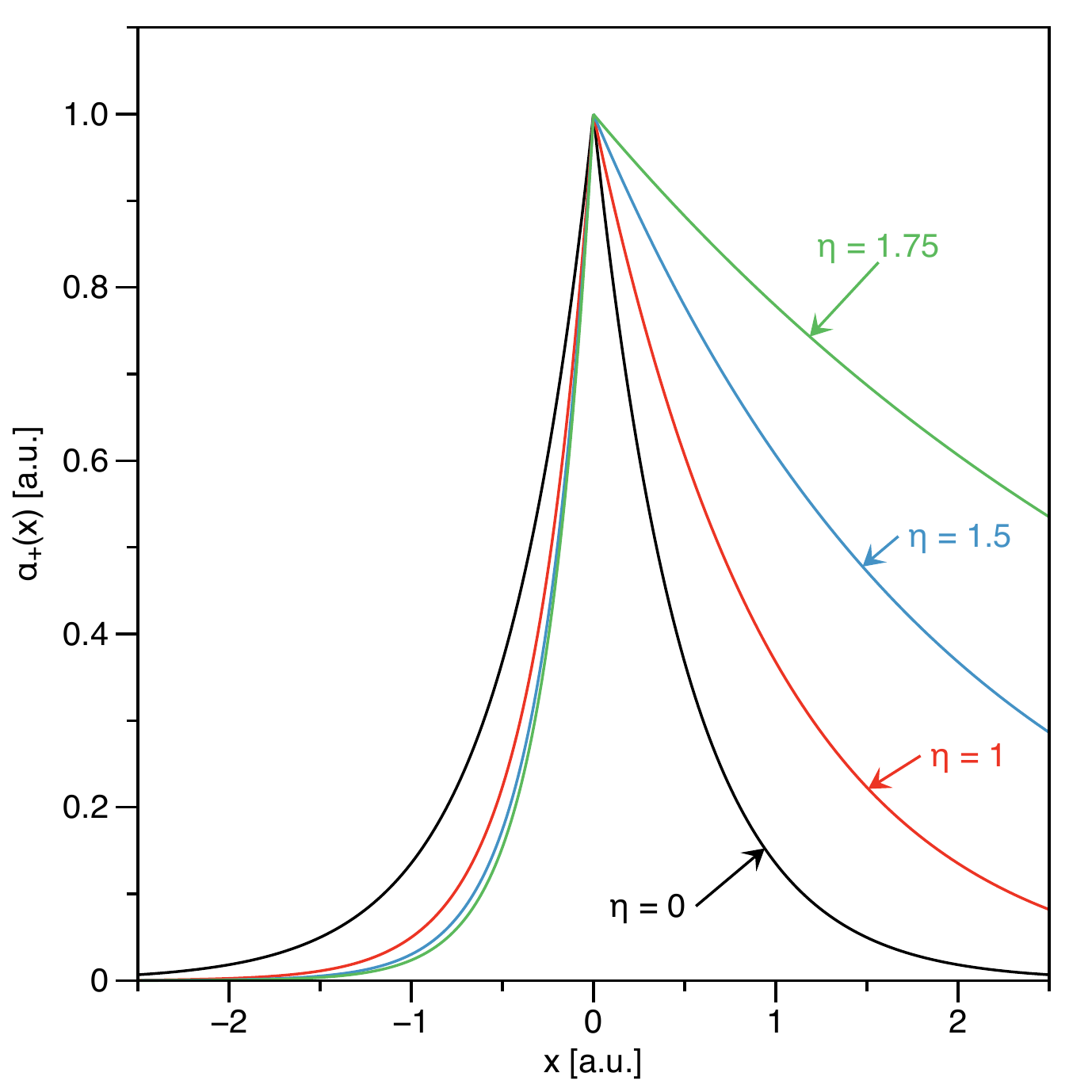}
  \caption{The orbital $\alpha_+(x)$ which is shifted to the right compared to the original orbital $\alpha(x)$ for several values of the correlation factor $\eta$ and with a nuclear charge $Z = 2$. Without correlation ($\eta = 0$) we have $\alpha_+(x) = \alpha(x)$. The other orbital $\alpha_-(x)$ is similar, though left-shifted.}
  \label{fig:coshOrb}
\end{figure}

In the harmonium atom and the hydrogen molecule, the transition to the strongly correlated regime was important to obtain some significant positive NO coefficients. Our simple Hylleraas Ansatz is not able to describe the strong correlation situation, so first we modify our model wavefunction such that we have a parameter to make a smooth transition into the strong correlation regime. By modifying the correlation function to $\cosh(\eta x_{12})$, the simple orbital product with $\alpha(x) = \e^{-Z\abs{x}}$ is actually able to describe strong correlation. The wavefunction now becomes
\begin{align}\label{eq:PsiCoshDef}
\Psi_{\cosh}(x_1,x_2) \isDefinedAs K\alpha(x_1)\alpha(x_2)\cosh(\eta x_{12}),
\end{align}
where $x_{12} \coloneqq x_1 - x_2$ and $K$ is a normalization constant which can be determined explicitly as
\begin{align*}
K = \frac{2Z(Z^2 - \eta^2)}{\sqrt{2(2Z^4 - 2Z^2\eta^2 + \eta^4)}}.
\end{align*}
By explicitly writing out the hyperbolic cosine, one readily finds that this wavefunction can alternatively be written as
\begin{align*}
\Psi_{\cosh}(x_1,x_2) = \frac{K}{2}\bigl(\alpha_+(x_1)\alpha_-(x_2) + \alpha_-(x_1)\alpha_+(x_2)\bigr),
\end{align*}
where $\alpha_{\pm}(x) \coloneqq \e^{-Z\abs{x}\pm\eta x}$. Hence, the correlation factor is strong enough to deform the orbital into a left and a right shifted orbital (see Fig.~\ref{fig:coshOrb}). We recognize the typical form of a strongly correlated Heitler--London type wave function~\cite{HeitlerLondon1927} as e.g.\ also appears in the dissociating H$_2$ molecule. Atomic wave functions of this form with left and right polarized orbitals have also been considered to describe correlated electrons in strong laser fields~\cite{DahlenLeeuwen2001, Dahlen2002}. Since only two orbitals are involved and the system is symmetric, the occupied NOs are readily constructed by making symmetry adapted combinations of the $\alpha_{\pm}(x)$ orbitals, $\phi_g(x) = C_g\e^{-Z\abs{x}}\cosh(\eta x)$ and $\phi_u(x) = C_u\e^{-Z\abs{x}}\sinh(\eta x)$, with normalization constants
\begin{align*}
\abs{C_g}^2 &= \frac{2Z(Z^2 - \eta^2)}{2Z^2 - \eta^2} &
&\text{and} &
\abs{C_u}^2 &= \frac{2Z(Z^2 - \eta^2)}{\eta^2}.
\end{align*}
The corresponding natural amplitudes are readily obtained by equating the spectral expansion with the original wavefunction, which gives
\begin{subequations}
\begin{align}
\label{eq:statcCoefEven}
c_g &= \hphantom{-}\frac{K}{C_g^2} = \frac{1}{\sqrt{2}}\frac{2Z^2 - \eta^2}{\sqrt{2Z^4 - 2Z^2\eta^2 + \eta^4}}, \\
\label{eq:statcCoefOdd}
c_u &= -\frac{K}{C_u^2} = \frac{1}{\sqrt{2}}\frac{-\eta^2}{\sqrt{2Z^4 - 2Z^2\eta^2 + \eta^4}}.
\end{align}
\end{subequations}
In the limit $\eta \to Z$ the amplitudes become $c_g = -c_u = 1/\sqrt{2}$, so this correlation function is indeed able to recover the strong correlation limit.

Though we have constructed a model wavefunction with a parameter $\eta$ that allows for a smooth transition between between a weakly and strongly correlated regime, we have only two occupied NOs. To make all the NOs fractionally occupied we add a cusp to the correlation function of the form $\xi\sinh(\eta\abs{x_{12}})$, where $\xi > 0$ is an additional parameter to vary the strength of the cusp. The full wavefunction now becomes
\begin{multline}\label{eq:coshsinhPsi}
\Psi_{\text{cs}}(x_1,x_2) = K\alpha(x_1)\alpha(x_2) \\
{} \times \bigl(\cosh(\eta x_{12}) + \xi\sinh(\eta\abs{x_{12}})\bigr).
\end{multline}
The NOs of this wavefunction can be obtained with similar methods as the NOs of the 1D Hylleraas function~\cite{GiesbertzLeeuwen2013a}. First we write the eigenvalue equation for the NOs as
\begin{multline*}
K\integ{x_2}\bigl(\cosh(\eta x_{12}) + \xi\sinh(\eta\abs{x_{12}})\bigr) \\
{} \times \alpha(x_2)^2\varphi_k(x_2)= c_k\,\varphi_k(x_1),
\end{multline*}
where $\varphi_k(x) \isDefinedAs \phi_k(x) / \alpha(x)$. By taking twice the derivative with respect to $x_1$, this integral equation can be turned into a differential equation for $\varphi_k(x)$
\begin{align}\label{eq:diffEqCoshSinh}
\varphi_k''(x) = \bigl(\lambda_k\alpha^2(x) + \eta^2\bigr)\varphi_k(x),
\end{align}
where $\lambda_k \isDefinedAs 2 \eta\xi K/ c_k$. Solving the differential equation is rather technical and has been deferred to the Appendix including its solutions. The most important result is that although we have a wavefunction which includes strong correlations, there is still only one positive NO coefficient. Hence, although multiple positive natural amplitudes have been considered a strong correlation effect~\cite{CioslowskiPernal2000b, CioslowskiPernal2006}, we find from our exactly solvable model system that the appearance of multiple positive natural amplitudes is actually \emph{not} caused by strong correlations. Strong correlation effects only seem to enhance their magnitude, but not to be essential for their existence.

If strong correlation effects are not the cause for the existence of multiple positive natural amplitudes, what property is essential for their existence? An important clue comes the fact that positive natural amplitudes are important to describe the Van der Waals effects in molecular dissociation~\cite{ShengMentelGritsenko2013}. The Van der Waals effects are caused by the long-range nature of the Coulomb interaction, so we expect some relation to the positivity of the NO coefficients. The $1/r$ tail of the Coulomb interaction causes the wavefunction not to decay merely exponentially as one electron is pulled away, but to behave asymptotically as $r^s\e^{-\sqrt{2 I_1}r}$, where $I_1$ is the first ionization energy and $s \isDefinedAs (Q - N + 1)/\sqrt{2I_1} - 1$, with $Q$ as the total nuclear charge and $N$ as the total number of electrons~\cite{KatrielDavidson1980}. The exponential decay is caused by the kinetic energy in combination with the required normalizability of the wavefunction. The additional factor $r^s$ is due to the $1/r$ tail of the Coulomb potential. To incorporate this asymptotic behavior, we change the wavefunction to
\begin{align*}
\Psi(x_1,x_2) = K\,\alpha(x_1)\alpha(x_2) f_{\text{pow}}(x_{12})
\end{align*}
with the correlation function
\begin{align}\label{eq:coshsinhxsCorFunc}
f_{\text{pow}}(x_{12}) \isDefinedAs (1 + \abs{x_{12}})^s\cosh(\eta x_{12}),
\end{align}
where we used $(1 + \abs{x_{12}})$ instead of only $x_{12}$ to ensure that the wavefunction does not vanish for $x_{12} \to 0$. Since the term $(1 + \abs{x_{12}})^s$ already introduces a cusp for $s \neq 0$, the hyperbolic sine is not needed anymore. Unfortunately, we were not able to find an analytic solution using this correlation function, even in conjunction with the simple Slater function for the orbital. However, a diagonal representation of the wavefunction can still be obtained numerically on a grid. Since the orbital is a simple exponential, we used Gauss--Laguerre quadrature for the integration. As parameters we used $Z = 2.0$, $\eta = 1.0$ and $s = 1.0$. There is no stringent reason for this particular choice of parameters, except that they satisfy $Z > \eta > 0$. The most significant natural amplitudes are shown in Table~\ref{tab:NOcoefs}. Indeed we find that due to the additional term $(1 + x_{12})^s$ in the correlation function, some additional amplitudes of ungerade NOs are now positive as well. The effect of the additional $x_{12}^s$ term in the wavefunction is that the orbitals become now effectively polarized in the direction of the interaction. For $s = 1$ this polarization effect can be readily worked out explicitly for $x_1 > x_2$ as
\begin{multline*}
\Psi(x_1 > x_2) = \frac{K}{2}\Bigl[\alpha_+(x_1)\alpha_-(x_2) + \alpha_-(x_1)\alpha_+(x_2) \\
{} + x_1\alpha_+(x_1)\alpha_-(x_2) + x_1\alpha_-(x_1)\alpha_+(x_2) \\
{} - \alpha_+(x_1)x_2\alpha_-(x_2) - \alpha_-(x_1)x_2\alpha_+(x_2)\Bigr],
\end{multline*}
where the orbitals $\alpha_{\pm}(x)$ have been introduced at the beginning of this subsection. We clearly see that the effect of the additional $x_{12}$ is to mix in polarized orbitals of the form $x\alpha_{\pm}(x)$. These additional $p$-orbitals represent exactly the induced dipole-dipole polarization used in the physical interpretation of the Van der Waals interaction~\cite{ShengMentelGritsenko2013}. The physical picture is therefore that if we place an electron at a large distance from the atomic centre this electron polarizes the remaining atom due to the long-range tail of the Coulomb interaction. The corresponding large distance structure of the wave function is responsible for the appearance of positive natural amplitudes.

\begin{table}[tb]
\caption{Numerical NO coefficients for a wavefunction $\Psi(x_1,x_2) = K\,\e^{-Z\abs{x_1}}\e^{-Z\abs{x_2}}f_{\text{pow}}(x_{12})$ with parameters $Z = 2.0$, $\eta = 1.0$ and $s = 1.0$. Only the numerically significant ($\abs{c_k} > 10^{-15}$) positive amplitudes are shown and the five largest negative NO coefficients in each irreducible representation.}
\label{tab:NOcoefs}
\begin{ruledtabular}
\begin{tabular}{d d  d d @{\rule{6.5ex}{0pt}}}
\multicolumn{2}{c}{gerade}				&\multicolumn{2}{c}{ungerade}				\\
\hline
\rule{0pt}{2.5ex}
9.463 \cdot 10^{-1}	&-5.916 \cdot 10^{-2}	&1.485 \cdot 10^{-3}		&-3.162 \cdot 10^{-1} \\
1.120 \cdot 10^{-5}	&-1.544 \cdot 10^{-2}	&1.506 \cdot 10^{-8} 	&-2.115 \cdot 10^{-2} \\
2.288 \cdot 10^{-11}	&-6.864 \cdot 10^{-3}	&3.050 \cdot 10^{-14}	&-7.615 \cdot 10^{-3} \\
				&-3.992 \cdot 10^{-3}	&					&-4.156 \cdot 10^{-3} \\
				&-2.770 \cdot 10^{-3}	&					&-2.787 \cdot 10^{-3} \\
				&\vdots				&					&\vdots
\end{tabular}	
\end{ruledtabular}
\end{table}

\section{How do the NOs change their sign?}

\begin{figure}[t]
  \includegraphics[width=\columnwidth]{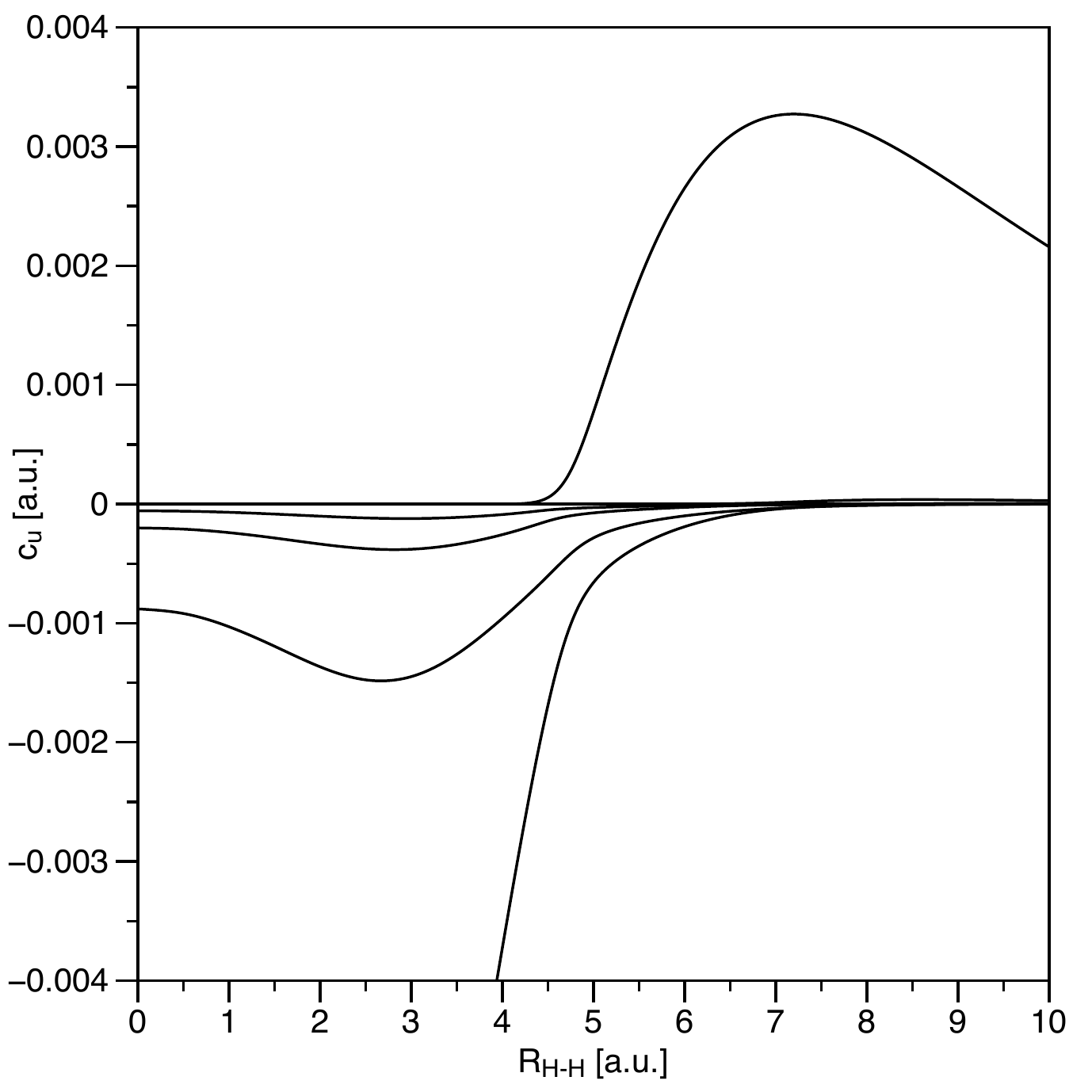}
  \caption{The coefficients in the NO expansion of the wavefunction of the 1D H$_2$ system for the odd NOs versus the distance, $R_{\text{H--H}}$,  between the two nuclei. The coefficient of the highest occupied ungerade NO is off the scale, so it is not visible in the plot.}
  \label{fig:cu}
\end{figure}

In the previous section we have shown that the existence of multiple positive natural amplitudes is caused by the long-range behavior of the Coulomb potential and not a feature of strong correlations. Therefore, positive NO coefficients should exist in all Coulomb systems and probably there is an infinite amount of them. Hence, no additional positive amplitudes need to be created when making the transition to a strongly correlated state, since they are simply already there, though very small. We have investigated this transition ourselves by performing calculations on a one dimensional (1D) hydrogen molecule with the Hamiltonian
\begin{multline*}
\hat{H} = -\half\frac{\du^2}{\du x_1^2} - \half\frac{\du^2}{\du x_2^2} \\
{} + v_{\text{ext}}(x_1) + v_{\text{ext}}(x_2) + v_{\text{soft}}(x_1-x_2),
\end{multline*}
where the external potential is defined as
\begin{align*}
v_{\text{ext}}(x) = v_{\text{soft}}\left(x - \frac{R_{\text{H--H}}}{2}\right) + v_{\text{soft}}\left(x + \frac{R_{\text{H--H}}}{2}\right).
\end{align*}
The advantage of reducing the problem to 1D is that the problem becomes computationally more tractable while still retaining the most important physics which is along the bond axis. Since the Coulomb potential is too singular in 1D to give finite energies, we have used soft Coulomb potentials
\begin{align*}
v_{\text{soft}}(x) \isDefinedAs \frac{1}{\sqrt{x^2 + 1}}
\end{align*}
instead. For this low dimensional system the Schrödinger equation can be solved directly on a grid. To reduce the computational cost, only the unique wedge in the centre-of-mass frame has been calculated on a $1600 \times 1600$ grid. The derivatives have been discretized by using cubic splines~\cite{PhD-Becke1981, Becke1982} and for the diagonalization we used the iterative Davidson algorithm~\cite{Davidson1975}. The wavefunction coefficients~\eqref{eq:PsiSpectral} were obtained by first transforming the wavefunction back to the normal coordinate frame $(x_1,x_2)$ and then the wavefunction was diagonalized with the \textsc{lapack} routines~\cite{LAPACK}.

The results for the coefficients corresponding to the ungerade NOs are shown in Fig.~\ref{fig:cu}. As expected, we clearly see that one of the NOs gains a rather large amplitude upon dissociation. However, this NO does not obtain a positive amplitude by crossing zero. Instead a large amount of avoided crossings occur around $R_{\text{H--H}} \approx 4$ Bohr to carry over the NO character to an already existing positive natural amplitude. In retrospect this result is not surprising, since we know that the eigenvalues of a matrix do not become degenerate in general if only one parameter is varied. Only under very special circumstances such a degeneracy can occur as is well known for conical intersections~\cite{VonNeumannWigner1929, Teller1937}. The argument goes as follows. Consider two NOs, $\phi_k$ and $\phi_l$ which are almost degenerate. A perturbation in the Hamiltonian, such as a change in the bond distance, will give rise to a new ground state $\Psi$ with new NOs and coefficients. Within the two-dimensional subspace spanned by the unperturbed orbitals $\phi_k$ and $\phi_l$, the new NO coefficients are readily determined to be
\begin{align*}
c_{\pm} = \frac{\Psi_{kk}+\Psi_{ll}}{2} \pm \sqrt{\left(\frac{\Psi_{kk}-\Psi_{ll}}{2}\right)^{\mathrlap{2}} + \abs{\Psi_{kl}}^2},
\end{align*}
where $\Psi_{ij} \isDefinedAs \brakket{\phi_i}{\hat{\Psi}}{\phi_j}$, where $\hat{\Psi}$ is the wavefunction regraded as an integral operator. The term under the square root gives the magnitude of the splitting between the natural amplitudes. Since the splitting consists of two positive contributions, the amplitudes can only be degenerate if they both vanish identically
\begin{align}\label{eq:degCond}
\Psi_{kk} - \Psi_{ll} = \Psi_{kl} = 0.
\end{align}
This condition is rarely met, since we have to satisfy two constraints with only one parameter for real wavefunctions~\cite{LandauLifshitz1977}, so the crossing is typically avoided as in Fig.~\ref{fig:cu}. This implies that a NO coefficient can never become zero, because there is an infinite amount of eigenvalues clustered around zero, hence there will always be an eigenvalue closer to zero which has to be crossed. Hence, the only general way an NO can obtain a coefficient with a different sign is to pass its character over to an other natural amplitude of opposite sign, via avoided crossings with all eigenvalues near zero, which is exactly what we observe in Fig.~\ref{fig:cu}.

It is still possible that an NO coefficient can change its sign by passing through zero if condition~\eqref{eq:degCond} is met, though it has to be satisfied for all the NOs involved, i.e.\ all the NO coefficients clustered around zero. Since condition~\eqref{eq:degCond} also implies that the splitting between all these NO coefficients vanishes, all their coefficients have to collapse collectively to zero. This is a rather exceptional situation which probably can never be achieved  by varying the external potential in a system with Coulomb interactions. We will show this collapse in a system with different interactions by varying the interaction strength later in this section. Of course, when a finite basis representation of the Hamiltonian is used, the NO with the smallest NO amplitude on the positive or negative side does not have an other NO closer to zero to prevent this coefficient from crossing zero as has been observed before~\cite{CioslowskiPernal2006, ShengMentelGritsenko2013}.

\begin{figure}[t]
  \includegraphics[width=\columnwidth]{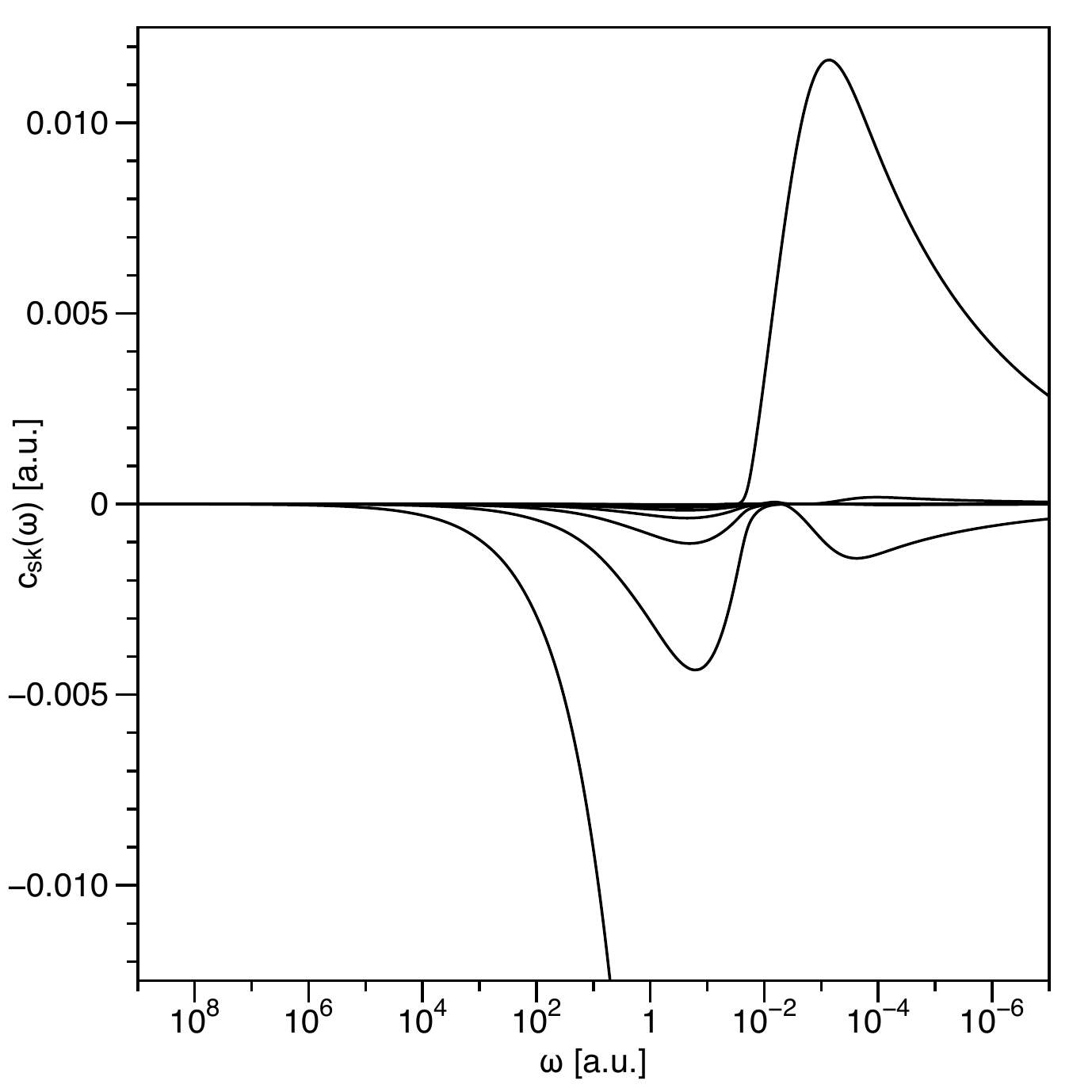}
  \caption{The coefficients in the NO expansion of the ground state of the harmonium atom in the $s$-channel as a function of the confinement $\omega$. The first amplitude is off the scale, so not visible in the plot.}
  \label{fig:harmoniumNOs}
\end{figure}

The harmonium atom uses the same mechanism to change effectively the sign of its NOs. In the original plots~\cite{CioslowskiPernal2000b, PhD-Pernal2002} this is not so clear since the occupation numbers are plotted on a log-log scale instead of the natural amplitudes directly. To generate the NO coefficients, we have implemented the numerical solution for the harmonium atom as described in Refs~\cite{CioslowskiPernal2000b, PhD-Pernal2002}. To obtain the NO coefficients, we used a different approach. The harmonium ground state wavefunction is of the form
\begin{align*}
\Psi(\vecr_1,\vecr_2) = N\e^{-\half\omega r_1^2}\e^{-\half\omega r_2^2}f(r_{12}),
\end{align*}
where $r_{12} \isDefinedAs \abs{\vecr_1 - \vecr_2}$ and $N$ is a normalization constant. Due to the spherical symmetry, the NOs are also eigenfunctions of the angular momentum operator and can be classified accordingly. The correlation function can be expanded in Legendre polynomials, $\LegendreP_l(s)$, as
\begin{align*}
f(r_{12}) = \sum_{\l=0}^{\infty}f_l(x_1,x_2)\LegendreP_l(s),
\end{align*}
where $s \isDefinedAs \cos\theta_{12}$, so the NOs are degenerate in each $l$-channel and can be calculated per $l$-channel separately. The Fourier coefficients of the correlation function can be determined as
\begin{multline*}
f_l(r_1,r_2)
= \frac{2l+1}{2}\binteg{s}{-1}{1}f\left(\sqrt{r_1^2 + r_2^2 -2r_1r_2s}\right)\LegendreP_l(s) \\
= \frac{2l+1}{2r_1r_2}\binteg{r_{12}}{\abs{r_1-r_2}}{r_1+r_2}r_{12}f\bigl(r_{12}\bigr)
\LegendreP_l\left(\frac{r_1^2 + r_2^2 - r_{12}^2}{2r_1r_2}\right).
\end{multline*}
The last form was used in practice, since the square-root impaired the accuracy of the numerical integration. Since the wavefunction is composed of a product of Gaussians, it is compelling to use Gauss--Hermite quadrature for the integration in the eigenvalue equation for the NO, so we have
\begin{align*}
\frac{4\pi N}{2l+1}\sum_j\e^{-\half\omega r_i^2}f_l(r_i,r_j)r_j^2w_j \cdot \phi_{lk}(r_j) = c_{lk}\,\phi_{lk}(r_i),
\end{align*}
where $w_i$ are the weights and $r_i$ are the abscissas. The NO eigenvalue equation has now been discretized on the Gauss--Hermite abscissas and has been turned into an effective non-symmetric matrix eigenvalue equation. This eigenvalue equation is trivially symmetrized by multiplying the equation with some suitable factors as
\begin{multline*}
\frac{4\pi N}{2l+1}\sum_j
\frac{r_i\sqrt{w_i}}{\e^{\frac{1}{4}\omega r_i^2}}f_l(r_i,r_j)\frac{r_j\sqrt{w_j}}{\e^{\frac{1}{4}\omega r_j^2}} \cdot
\frac{r_j\sqrt{w_j}}{\e^{-\frac{1}{4}\omega r_j^2}}\phi_{lk}(r_j) \\
= c_{lk}\,\frac{r_i\sqrt{w_i}}{\e^{-\frac{1}{4}\omega r_i^2}}\phi_{lk}(r_i).
\end{multline*}
Results for the natural amplitudes in the $s$-channel are shown in Fig.~\ref{fig:harmoniumNOs}. We have reverted the $\omega$ axis, so that the strong correlation regime is on the right, as is also the case for the plot of the NO coefficients of the odd NOs of 1D hydrogen molecule (Fig.~\ref{fig:cu}). The behavior is roughly the same as in the H$_2$ case: when going into the strong correlation regime, one of the positive coefficients picks up a large amplitude (maximum at $\omega \approx 10^{-3}$) and decays slowly again when progressing further into the strongly correlated regime (decreasing $\omega$). It is also clear that this coefficient was not originally negative, but gains its large amplitude from an NO that initially had a significant negative natural amplitude and carries over its character to this positive NO coefficient by an infinite amount of avoided crossings with all the natural amplitudes located around zero. Exactly as in the 1D H$_2$ case, this is the only way how the NOs (eigenfunctions of the wavefunction) could change their sign, since their amplitudes cannot become zero due to the infinite amount of eigenvalues piled up around zero.

That the natural amplitude cannot vanish due to the infinite clustering around zero, is also a very strong argument in favor of the statement that there are no zero occupation numbers in Coulomb systems, since the same argument applies to the occupation numbers of the 1RDM. Because there is an infinite amount of occupation numbers close to zero, there will always be an occupation number closer to zero which prevents occupation numbers to become zero by making a perturbation to the system, i.e.\ changing the external potential. The only possibility is that an occupation number is already zero, but the same argument implies that this occupation number can never become nonzero, since there is always an occupation number smaller than the value we want to give this occupation number. Since we will need a complete set of NOs to expand the non-analytic behavior of the 1RDM, it would be strange if some of them can be left out irrespective of any perturbation of the system. Therefore, all occupation numbers should always be non-zero in systems with a Coulomb interaction.

\begin{figure}[t]
  \includegraphics[width=\columnwidth]{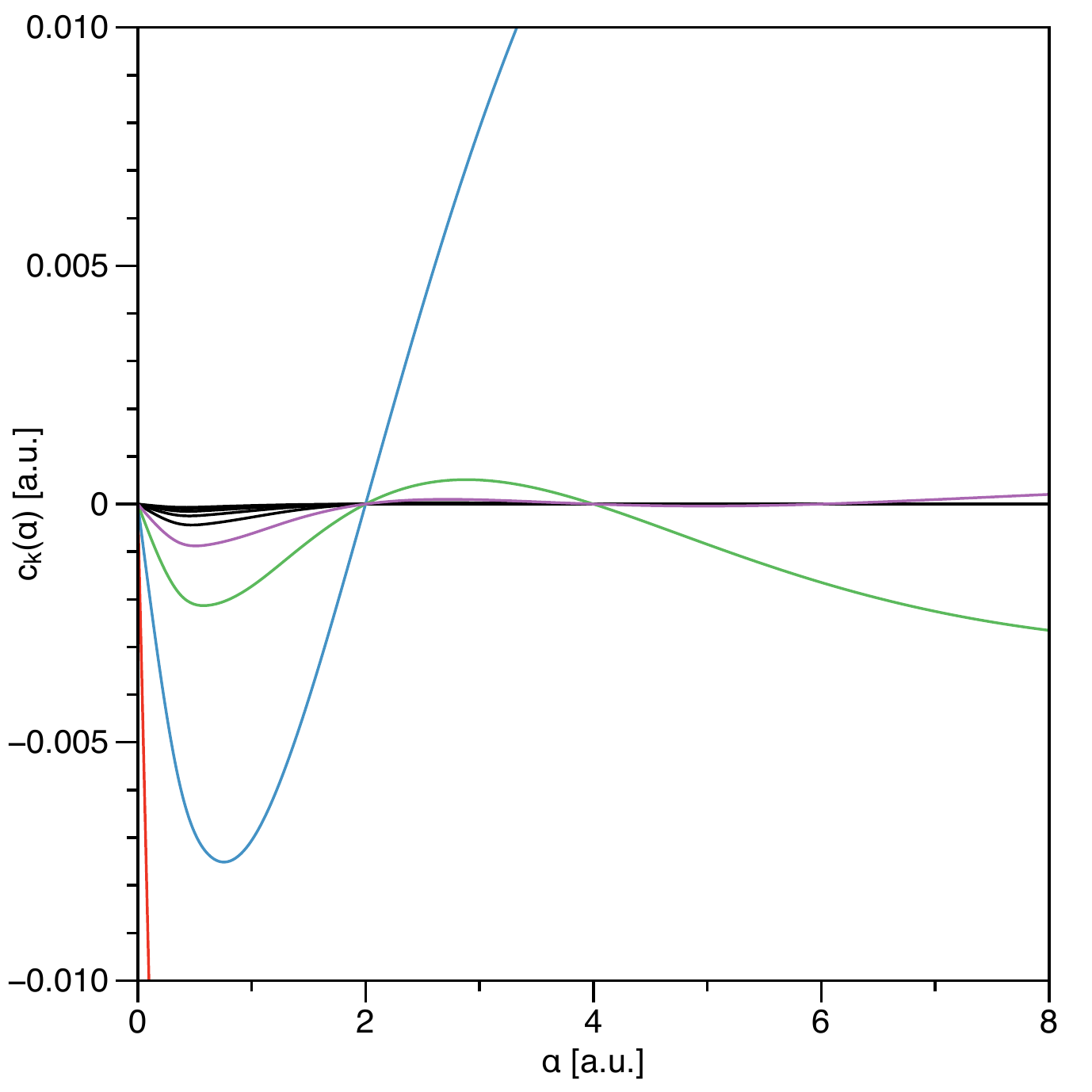}
  \caption{The coefficients in the NO expansion of the ground state of the system with inverse harmonic interactions in the $s$-channel as a function of $\alpha$. The strength of the external potential has been set to $\omega = 1$. The first NO coefficient is off the scale, so not visible in the plot.}
  \label{fig:invHarmNOs}
\end{figure}

As mentioned before, only under very special circumstances the eigenvalues of a matrix can become degenerate under the variation of one parameter. Since there are an infinite amount of occupation numbers close to zero, we will need to make all these eigenvalues zero to facilitate a true sign change of an expansion coefficient. Such a massive degeneracy can only be created by modifying the interaction in a very particular way. As an example consider the system with inverse harmonic interactions with the Hamiltonian
\begin{align*}
\hat{H} = -\half\nabla_{\vecr_1}^2 - \half\nabla_{\vecr_2}^2 +
\half\omega^2(r_1^2 + r_2^2) + \frac{\lambda}{r_{12}^2}.
\end{align*}
This system has been considered before by Morrison et al.\ to study the existence of unoccupied NOs~\cite{MorrisonZhouParr1993}. The ground state of this Hamiltonian has the following surprisingly simple form~\cite{MorrisonZhouParr1993, LandauLifshitz1977}
\begin{align*}
\Psi(\vecr_1,\vecr_2) = \sqrt{\frac{\omega^{3+\alpha}}{2^{1+\alpha}\pi^{5/2}\Gamma\bigl(\frac{3}{2}+\alpha\bigr)}}\e^{-\half\omega(r_1^2+r_2^2)}r_{12}^{\alpha},
\end{align*}
where $\alpha = \bigl(\sqrt{1+4\lambda}-1\bigr)/2$~\footnote{The factor $1/2$ in the exponent seems to be missing in Ref.~\cite{MorrisonZhouParr1993}. This error propagates throughout the article.}. This ground state wavefunction has the special property that for even $\alpha$ it becomes separable, i.e.\ the wavefunction can be exactly expanded in a finite number of orbitals. In Ref.~\cite{GiesbertzLeeuwen2013a} we have shown that such an expansion is only possible for even $\alpha$ and proved that for all other $\alpha$, all of the occupation numbers are non-zero. Therefore, if we vary the interaction strength, so effectively $\alpha$, the spectrum of almost all NO coefficients should collapse to zero at even $\alpha$. In Fig.~\ref{fig:invHarmNOs} we show the NO coefficients in the $s$-channel and indeed at even $\alpha$ almost all the amplitudes collapse to zero and a true sign change of the natural amplitudes is possible at these points. As expected, after each crossing an additional NO remains non-zero at the next crossing, since at each next crossing an additional NO is required for the expansion of $r_{12}^{\alpha}$. The natural amplitudes in the higher $l$-channels show similar features, except that their first finite occupancy is only required from $\alpha = 2l$ onwards.

This massive degeneracy at zero could only be achieved by replacing the Coulomb interaction by a $1/r^2$ interaction and making variations in the interaction, so not by variations in the external potential. It is therefore expected, that such a massive degeneracy can never be achieved in a system with Coulomb interactions.

\section{Conclusion}
We have shown that there are always an infinite amount of positive NO coefficients present in systems with a Coulomb interaction. The presence of these positive coefficients is caused by fractional power in the decay of the wavefunction, $r^s\e^{-\sqrt{2I_1}r}$, due to the long-range behavior of the Coulomb potential. In the weakly correlated regime only one of the positive NO coefficients makes a significant contribution. The other positive NO coefficients only play a noteworthy role when the long-range character of the Coulomb interaction become important, i.e.\ in the strong correlation regime. The foremost example is the Van der Waals interactions in a dissociating hydrogen molecule.

Earlier studies by Cioslowski and Pernal~\cite{CioslowskiPernal2000b, CioslowskiPernal2006} assumed that only one positive natural amplitude exists in the weakly correlated regime, so they needed the amplitudes to cross zero to explain the transition to the alternating sign pattern of the NO coefficients in the strong correlation limit. Due to our observation that there are actually always an infinite amount of positive natural amplitudes, even in the weakly correlated regime, this argument does not hold anymore. Actually, the natural amplitudes themselves never change sign, but the character of the NOs is carries over to different amplitudes with opposite sign when making the transition to the strongly correlated regime. This explanation for the apparent sign change of the NOs has been confirmed by numerical calculations on the harmonium atom and the 1D hydrogen molecule.

The mechanism of avoided crossings also prevents the NO coefficients from becoming zero by making variations in the external potential, e.g.\ nuclear configuration of a molecule. Since there is an infinite amount of natural amplitudes clustered around zero, there will always be an amplitude closer to zero than the amplitude we would like to make zero by varying the external potential. Because there can only be an avoided crossing, the natural amplitudes can never cross each other to reach zero. So effectively the infinite amount of natural amplitudes close to zero act as a wall preventing other amplitudes from becoming zero when the external potential is varied. The only way a NO coefficient can become zero is by making special variations in the interaction itself, such that the crossing will not be avoided anymore. In such a situation all the NO coefficients near zero will collapse collectively to zero as has been demonstrated in a system with scalable inverse harmonic interactions, $\lambda/r_{12}^2$. However, this situation can never be achieved in Coulomb systems by merely varying the external one-body potential. The remaining alternative for the existence of vanishing amplitudes seems to be an NO amplitude which is always zero irrespective of any external perturbation. However, the existence of such a robust unoccupied NO does not seem to be very likely.

The same line of reasoning immediately carries over to the natural occupation numbers, so we have a very strong argument that the occupation numbers in systems with a Coulomb interaction never vanish. Note that this argument only applies to an infinite Hilbert space, so is not applicable to the model Hamiltonians due to the finite basis representation on our computers. In a finite Hilbert space there will be a smallest occupation number which can vanish, since there is no lower occupied NO anymore that can prevent this, via an avoided crossing.

\begin{acknowledgments}
The authors acknowledge the Academy of Finland for research funding under Grant No.\ 127739. KJHG also gratefully acknowledges a VENI grant by the Netherlands Foundation for Research NWO (722.012.013).
\end{acknowledgments}

\appendix*

\section{NOs of the cosh-sinh correlated wavefunction}
\label{ap:coshsinhNOs}
The steps to calculate the NOs of the wavefunction~\eqref{eq:coshsinhPsi} are analogue to the procedure to calculate the NOs and coefficients exposed in Appendix~A of Ref.~\cite{GiesbertzLeeuwen2013a}. First we will determine the boundary conditions that the solutions $\varphi(x)$ need to satisfy. First we rewrite the integral equation as
\begin{multline*}
\frac{c_k}{K}\varphi_k(x) = \frac{1+\xi}{2} e^{\eta x}\binteg{y}{-\infty}{x}\e^{-\eta y}\alpha^2(y)\varphi_k(y) \\*
{} + \frac{1 - \xi}{2} e^{-\eta x}\binteg{y}{-\infty}{x}\e^{\eta y}\alpha^2(y)\varphi_k(y) \\*
{} + \frac{1+\xi}{2} e^{-\eta x}\binteg{y}{x}{\infty}\e^{\eta y}\alpha^2(y)\varphi_k(y) \\*
{} + \frac{1 - \xi}{2} e^{\eta x}\binteg{y}{x}{\infty}\e^{-\eta y}\alpha^2(y)\varphi_k(y).
\end{multline*}
In the limit $x \to \infty$ the last two integrals do not contribute, since the orbital $\alpha(x) = \e^{-Z\abs{x}}$ decays fast enough for large $x$, and we find for the large $x$ behavior
\begin{multline*}
\frac{c_k}{K}\varphi_k(x) = \frac{1+\xi}{2}\e^{\eta x}\binteg{y}{-\infty}{\infty}\e^{-\eta y}w(y)\varphi_k(y) \\*
{} + \frac{1-\xi}{2}\e^{-\eta x}\binteg{y}{-\infty}{\infty}\e^{\eta y}w(y)\varphi_k(y).
\end{multline*}
Since the problem is symmetric, the solutions can be separated in the even and odd solutions. For the even solutions the integrals give the same value, so we find that the gerade solutions should behave asymptotically as
\begin{align*}
\varphi_{g}(x \to \infty) &\sim \e^{\eta x} + \frac{1-\xi}{1+\xi}\e^{-\eta x}.
\end{align*}
In the case of the odd solutions, the integrals are minus each other, so we find that the odd solutions must behave asymptotically as
\begin{align*}
\varphi_{u}(x \to \infty) &\sim \e^{\eta x} - \frac{1-\xi}{1+\xi}\e^{-\eta x}.
\end{align*}
This asymptotic analysis only assumes that $\alpha^2(x)$ is symmetric and decays sufficiently fast.

Since we are looking for symmetry adapted solutions, we only need to solve the problem for $x \geq 0$. If we introduce the following new variable
\begin{align}\label{eq:sDef}
s(x) = \frac{\sqrt{\abs{\lambda}}}{Z}\e^{-Z x},
\end{align}
then for a function $\varphi(x) = f\bigl(s(x)\bigr)$ we have
\begin{align*}
\frac{\ud \varphi}{\ud x} &= \frac{\ud f}{\ud s}\frac{\ud s}{\ud x} = -\sqrt{\abs{\lambda}}\e^{-Zx}f'(s) = -Zs f'(s) \\
\frac{\ud^2\varphi}{\ud x^2} &= f''(s)\left(\frac{\ud s}{\ud x}\right)^{\mathrlap{2}} + f'(s)\frac{\ud^2 s}{\ud x^2} \\
&= Z^2\bigl(s^2f''(s) + sf'(s)\bigr),
\end{align*}
so the differential equation in terms of $f(s)$ becomes
\begin{align*}
s^2\frac{\ud^2 f}{\ud s^2} + s\frac{\ud f}{\ud s} - \left(\sgn(\lambda)s^2 + \frac{\eta^2}{Z^2}\right)f = 0.
\end{align*}
For $\lambda < 0$ the solutions are $\BesselJ_{\pm\nu}(s)$ with $\nu = \abs{\eta/Z}$. These solutions are only independent for $\nu \notin \Nats$. However, this only occurs for $\eta = 0$, in which case we reduce to the non-interacting solution. Note that situations like $\eta = Z$ cannot occur, since in those cases the wavefunction is not normalizable anymore, i.e. an unbound state. Imposing the boundary conditions at the origin, we find for the odd and even solutions respectively
\begin{align*}
\varphi_u(x) &= C\left[\BesselJ_{-\nu}\bigl(\tilde{\lambda}\bigr)\BesselJ_{\nu}\bigl(s(x)\bigr) - 
\BesselJ_{\nu}\bigl(\tilde{\lambda}\bigr)\BesselJ_{-\nu}\bigl(s(x)\bigr)\right], \\
\varphi_g(x) &= C\left[\BesselJ'_{-\nu}\bigl(\tilde{\lambda}\bigr)\BesselJ_{\nu}\bigl(s(x)\bigr) - 
\BesselJ'_{\nu}\bigl(\tilde{\lambda}\bigr)\BesselJ_{-\nu}\bigl(s(x)\bigr)\right],
\end{align*}
where $\tilde{\lambda} \isDefinedAs \sqrt{\abs{\lambda}}/Z$ and $\nu \isDefinedAs \eta/Z$. Imposing the boundary conditions at the other end ($x\to\infty$), we will obtain the quantization of $\tilde{\lambda}$. From the asymptotic behavior of the Bessel functions for small $s$, we find that for $x \to \infty$
\begin{align}\label{eq:BesselLimit}
\BesselJ_{\pm\nu}\Bigl(\tilde{\lambda}\e^{-Z x}\Bigr) 
\sim \frac{1}{\Gamma(1\pm\nu)}\biggl(\frac{\tilde{\lambda}}{2}\biggr)^{\pm\nu}\e^{\mp\abs{\eta}x},
\end{align}
where we assumed $\eta > 0$ without loss of generality. Hence, we find that for the odd solutions to have the correct asymptotic behavior, $\tilde{\lambda}$ needs to satisfy
\begin{align}\label{eq:minOddQuantization}
\frac{\BesselJ_{-\nu}\bigl(\tilde{\lambda}^-_u\bigr)}{\BesselJ_{\nu}\bigl(\tilde{\lambda}^-_u\bigr)}
\biggl(\frac{\tilde{\lambda}^-_u}{2}\biggr)^{\mathrlap{2\nu}} = R_{\nu,\xi},
\end{align}
where we introduced
\begin{align*}
R_{\nu,\xi} \coloneqq \frac{1-\xi}{1+\xi}\cdot\frac{\Gamma(1+\nu)}{\Gamma(1-\nu)}.
\end{align*}
Using the solutions $\tilde{\lambda}^-_{u,k}$ we can readily construct the ungerade NOs with negative coefficients as
\begin{multline*}
\phi^-_{u,k}(x) = C_{u,k}^-\,\sgn(x)\e^{-Z\abs{x}}  \\*
{} \times \biggl[\frac{\Gamma(1+\nu)}{1+\xi}\biggl(\frac{\tilde{\lambda}^-_{u,k}}{2}\biggr)^{\mathrlap{\nu}}
\BesselJ_{\nu}\Bigl(\tilde{\lambda}^-_{u,k}\e^{-Z\abs{x}}\Bigr) \\*
{} - \frac{\Gamma(1-\nu)}{1-\xi}\biggl(\frac{\tilde{\lambda}^-_{u,k}}{2}\biggr)^{\mathrlap{\nu}}
\BesselJ_{-\nu}\Bigl(\tilde{\lambda}^-_{u,k}\e^{-Z\abs{x}}\Bigr)\biggr],
\end{multline*}
where $C_{u,k}^-$ is a normalization constant. Similarly, for the even solutions we find the quantization condition
\begin{align}\label{eq:minEvenQuantization}
\frac{\BesselJ'_{-\nu}\bigl(\tilde{\lambda}^-_g\bigr)}{\BesselJ'_{\nu}\bigl(\tilde{\lambda}^-_g\bigr)}
\biggl(\frac{\tilde{\lambda}^-_g}{2}\biggr)^{\mathrlap{2\nu}} = -R_{\nu,\xi},
\end{align}
The corresponding NOs are readily constructed as
\begin{multline*}
\phi^-_{g,k}(x) = C_{g,k}^-\,\e^{-Z\abs{x}} \\*
{} \times \biggl[\frac{\Gamma(1+\nu)}{1+\xi}\biggl(\frac{\tilde{\lambda}^-_{g,k}}{2}\biggr)^{\mathrlap{\nu}}
\BesselJ_{\nu}\Bigl(\tilde{\lambda}^-_{g,k}\e^{-Z\abs{x}}\Bigr) \\*
{} + \frac{\Gamma(1-\nu)}{1-\xi}\biggl(\frac{\tilde{\lambda}^-_{g,k}}{2}\biggr)^{\mathrlap{\nu}}
\BesselJ_{-\nu}\Bigl(\tilde{\lambda}^-_{g,k}\e^{-Z\abs{x}}\Bigr)\biggr],
\end{multline*}
where $C_{g,k}^-$ is a normalization constant.

For the positive amplitudes ($\lambda > 0$), solutions are modified Bessel functions which can combined into odd an even combinations
\begin{align*}
\varphi_u(x) &= C\left[\BesselI_{-\nu}\bigl(\tilde{\lambda}\bigr)\BesselI_{\nu}\bigl(s(x)\bigr) - 
\BesselI_{\nu}\bigl(\tilde{\lambda}\bigr)\BesselI_{-\nu}\bigl(s(x)\bigr)\right], \\
\varphi_g(x) &= C\left[\BesselI'_{-\nu}\bigl(\tilde{\lambda}\bigr)\BesselI_{\nu}\bigl(s(x)\bigr) - 
\BesselI'_{\nu}\bigl(\tilde{\lambda}\bigr)\BesselI_{-\nu}\bigl(s(x)\bigr)\right].
\end{align*}
The modified Bessel functions of the first kind have the same small $s$ behavior as the `normal' Bessel functions of the first kind, so we immediately find
\begin{align*}
\frac{\BesselI_{-\nu}\bigl(\tilde{\lambda}\bigr)}{\BesselI_{\nu}\bigl(\tilde{\lambda}\bigr)}
\biggl(\frac{\tilde{\lambda}}{2}\biggr)^{\mathrlap{2\nu}} = R_{\nu,\xi}
\end{align*}
as a quantization condition for the odd solutions. The odd solution can be discarded however, since for $0 <  \eta < 1$ we have
\begin{align*}
\frac{\BesselI_{-\nu}\bigl(\tilde{\lambda}\bigr)}{\BesselI_{\nu}\bigl(\tilde{\lambda}\bigr)}
\biggl(\frac{\tilde{\lambda}}{2}\biggr)^{\mathrlap{2\nu}} > 1 > R_{\nu,\xi}
\end{align*}
on $0 < \nu < 1$ and $\xi > 0$, so odd positive solutions do not exist. For the even solutions, $\tilde{\lambda}$ needs to satisfy
\begin{align*}
\frac{\BesselI'_{-\nu}\bigl(\tilde{\lambda}^+_g\bigr)}{\BesselI'_{\nu}\bigl(\tilde{\lambda}^+_g\bigr)}
\biggl(\frac{\tilde{\lambda}^+_g}{2}\biggr)^{\mathrlap{\nu}} = -R_{\nu,\xi}.
\end{align*}
The left-hand side of the quantization condition at $\tilde{\lambda} = 0$ equals $\Gamma(\nu)/\Gamma(-\nu) > R_{\nu,\xi}$ with the inequality valid on $0 < \nu < 1$ and $\xi > 0$. Further, the left-hand side is a monotonically increasing function of $\tilde{\lambda}$, so for the physically relevant parameter range, $0 < \nu < 1$ and $\xi > 0$, there is exactly one positive solution. The corresponding gerade NO is readily constructed as
\begin{multline*}
\phi^+_g(x) = C_g^+\,\e^{-Z\abs{x}} \\*
{} \times \biggl[\frac{\Gamma(1+\nu)}{1+\xi}\biggl(\frac{\tilde{\lambda}^+_g}{2}\biggr)^{\mathrlap{-\nu}}
\BesselI_{\nu}\Bigl(\tilde{\lambda}^+_g\e^{-Z\abs{x}}\Bigr) \\*
{} + \frac{\Gamma(1-\nu)}{1-\xi}\biggl(\frac{\tilde{\lambda}^+_g}{2}\biggr)^{\mathrlap{\nu}}
\BesselI_{-\nu}\Bigl(\tilde{\lambda}^+_g\e^{-Z\abs{x}}\Bigr)\biggr],
\end{multline*}
As a check we consider the limit $\xi \to 0$ for which we should obtain only two occupied NOs as at the beginning of Sec.~\ref{sec:posCoefs}. First note that the NO coefficients are obtained as $c_k = 2 \eta\xi K/ \lambda_k$, so the only non-vanishing amplitudes can come from $\lambda_k \to 0$ and all the other $\lambda_k$ solutions give $c_k = 0$. To find these solutions, we make expansions of the quantization conditions around $\tilde{\lambda}$. Let us first start with the quantization condition for the negative odd solutions~\eqref{eq:minOddQuantization}. For the left-hand side we have
\begin{align*}
\frac{\BesselJ_{-\nu}\bigl(\tilde{\lambda}\bigr)}{\BesselJ_{\nu}\bigl(\tilde{\lambda}\bigr)}
\biggl(\frac{\tilde{\lambda}}{2}\biggr)^{\mathrlap{2\nu}}
= \frac{\Gamma(1 + \nu)}{\Gamma(1 - \nu)}\left(1 - \frac{\nu}{2(1 - \nu^2)}\tilde{\lambda}^2 + 
\Order\bigl(\tilde{\lambda}^4\bigr)\right).
\end{align*}
Since we now have only a linear equation in $\tilde{\lambda}^2$, the small $\lambda$ solutions is readily found as
\begin{align*}
\lambda = -Z^2\tilde{\lambda}^2 = -\frac{4\xi}{1+\xi}\frac{1 - \nu^2}{\nu}Z^2 < 0,
\end{align*}
for $0 < \nu < 1$. Calculating the corresponding amplitude and taking the limit $\xi \to 0$, we obtain indeed exactly the value of $c_u$ given in~\eqref{eq:statcCoefOdd}. Similarly, for the gerade quantization condition we find for small $\tilde{\lambda}$
\begin{multline*}
\frac{\BesselJ'_{-\nu}\bigl(\tilde{\lambda}\bigr)}{\BesselJ'_{\nu}\bigl(\tilde{\lambda}\bigr)}
\biggl(\frac{\tilde{\lambda}}{2}\biggr)^{2\nu} \\
= -\frac{\Gamma(1 + \nu)}{\Gamma(1 - \nu)}\left(1 + \frac{2 - \nu^2}{2\nu(1 - \nu^2)}\tilde{\lambda}^2 + 
\Order\bigl(\tilde{\lambda}^4\bigr)\right).
\end{multline*}
Now trying to solve the quantization condition~\eqref{eq:minEvenQuantization} for small $\tilde{\lambda}$ gives
\begin{align}\label{eq:geradeSmallLambda}
\lambda = -Z^2\tilde{\lambda}^2 = \frac{4\xi}{1 + \xi}\frac{\nu(1-\nu^2)}{2 - \nu^2}Z^2 > 0,
\end{align}
for $0 < \nu < 1$. Since this solution corresponds to a positive $\lambda$, we find that no finite negative NO coefficient exists for the gerade NOs.

Now we will investigate positive solutions. Since the modified Bessel functions are simply Bessel functions with an imaginary argument, only the sign in front of the $\tilde{\lambda}^2$ changes in the small $\tilde{\lambda}$ expansions. Because also the sign of $\lambda$ changes ($\lambda = Z^2\tilde{\lambda}^2$), the previously found small $\lambda$ solutions are also valid for positive $\lambda$. Since only the gerade solution has $\lambda > 0$, there remains one gerade NO with a positive coefficient and working out the corresponding natural amplitude with~\eqref{eq:geradeSmallLambda}, we correctly recover~\eqref{eq:statcCoefEven}. Further, using~\eqref{eq:BesselLimit} for the small $\tilde{\lambda}$ behavior of the Bessel functions in the NOs, the two NOs in the $\xi \to 0$ limit are recovered as well.

\bibliography{bible}

\end{document}